\documentclass[letterpaper, 10 pt, conference]{ieeeconf}  

\IEEEoverridecommandlockouts                              
\usepackage{amsmath,amssymb,amsthm,amsfonts,mathtools}
\usepackage{cite}
\usepackage{algorithmic}
\usepackage{graphicx}
\usepackage{bm}
\usepackage{multicol}
\usepackage{float}
\usepackage{xcolor}
\usepackage{subfigure}
\usepackage{textcomp}
\usepackage{url}
\let\labelindent\relax
\usepackage[inline]{enumitem}

\newtheorem{theorem}{Theorem}

\title{\LARGE \bf A Lie-Theoretic Approach to Propagating Uncertainty Jointly in Attitude and Angular Momentum}
\author{Amitesh S. Jayaraman, Jikai Ye, Gregory S. Chirikjian
\thanks{This work was supported by NUS Startup  grants A-0009059-02-00 and A-0009059-03-00, National Research Foundation, Singapore, under its Medium Sized Centre Programme - Centre for Advanced Robotics Technology Innovation (CARTIN),  sub award A-0009428-08-00, and AME Programmatic Fund Project MARIO A-0008449-01-00.}
\thanks{A. S. Jayaraman is with the Department of Mechanical Engineering, Stanford University, CA 94305, USA. Email: {\tt\small amiteshs@stanford.edu}} 
\thanks{J. Ye and G. S. Chirikjian are with the Department of Mechanical Engineering, National University of Singapore, Singapore. Email: {\tt\small \{jikai.ye, mpegre\}@nus.edu.sg}}
}

\begin{document}

\maketitle
\thispagestyle{empty}
\pagestyle{empty}

\begin{abstract}

Dynamic state estimation, as opposed to kinematic state estimation, seeks to estimate not only the orientation of a rigid body but also its angular velocity, through Euler's equations of rotational motion.
This paper demonstrates that the dynamic state estimation problem can be reformulated as estimating a probability distribution on a Lie group defined on phase space (the product space of rotation and angular momentum). 
The propagation equations are derived non-parametrically for the mean and covariance of the distribution.
It is also shown that the equations can be approximately solved by ignoring the third and higher moments of the probability distribution.
Numerical experiments show that the distribution constructed from the propagated mean and covariance fits the sample data better than an extended Kalman filter.

\end{abstract}


\section{Introduction} \label{sec:intro}
The field of attitude estimation was initially motivated by problems in the satellite and navigation community.
This led to the development of a variety of algorithms for state estimation and improvements to sensing modalities  \cite{markley2014fundamentals,zou2018algorithm,vedder1993star}. 
Recently, attitude estimation problems, often coupled with linear velocity estimation, have experienced renewed interest in the UAV/MAV community \cite{svacha2018inertial,svacha2019inertial,svacha2020imu}, and even in the jumping robot community \cite{kim2019}. 

Some papers such as \cite{solo2010attitude,fiori2019model,barrau2014intrinsic,shuster1993survey} formulate the problem in a Lie group setting. 
Such a coordinate-free approach is adopted in this paper, in contrast to \begin{enumerate*}[label=\itshape\alph*\upshape)] \item quaternion-based, \item direction-cosine matrix or \item Euler-angle based \end{enumerate*}  approaches. 
The first two approaches require additional renormalization and orthogonalization steps, respectively, to ensure the estimate describes a rotation.
The third approach has a well-known singularity in its description of rotations. 
A parameter-independent Lie-theoretic approach would respect the geometry of the configuration space, and has been widely used in nonlinear filtering applications \cite{brockett1973lie,brockett1976nonlinear,brockett1981nonlinear,barrau2018invariant}. 
Furthermore, it avoids the ambiguity in the choice of coordinates used to design conventional filters.

Given the interest in quadrotors with on-board inertial measurement units (IMUs), it is possible to estimate the attitude (orientation) and angular momentum together.
When inertial effects are small, one can ignore dynamics and cast the problem as a \emph{kinematic} error propagation. 
This has been explored in previous work, such as \cite{barfoot2014associating} and \cite{wang2008nonparametric}. 
When inertial effects are significant, one needs to make use of the (nonlinear) Euler equations of rotational motion to propagate the error in both orientation and angular momentum \emph{dynamically}. 
In \cite{bonnabel2020mathematical,brossard2021associating}, the authors consider a form of dynamic error propagation by propagating joint errors in position, orientation, and linear velocity on the extended pose group. 
Some work in the spacecraft attitude and rate estimation such as \cite{fujikawa1995spacecraft,sanyal2008global,ma2014magnetometer,psiaki2009generalized,palermo2001angular,oshman1999spacecraft,axelrad1996spacecraft} have considered variants of this problem.
In \cite{fujikawa1995spacecraft} and \cite{axelrad1996spacecraft}, the authors employ an extended Kalman filter to estimate the orientation and angular velocity in parameters (Euler angles/quaternion). 
In \cite{sanyal2008global}, the authors exclude noise in the process dynamics entirely and propagated measurement noise to obtain bounds for attitude and angular velocity estimates. 
Some papers, such as \cite{palermo2001angular} and \cite{psiaki2009generalized}, have sought to improve beyond traditional Kalman filter methods, where the latter frames the problem as a least-squares minimization over quaternion and angular velocity space subject to the unit quaternion constraint. 
Oshman and Markley \cite{oshman1999spacecraft} avoid using the Euler equations of rotational dynamics completely and instead model angular acceleration as a linear stochastic process, but this obscures the connection with the physical force balance and leads to additional tuning parameters for the filter. 
Our paper develops a non-parametric theory that propagates uncertainty of the orientation and angular momentum of a rigid body subject to viscous and stochastic disturbance torques. 

\textit{Contributions}: We cast the combined orientation and angular momentum uncertainty propagation problem non-parametrically on a group $SO(3)^T\ltimes\mathbb{R}^3$ \cite{jayaraman2023inertial}.
Furthermore, we also consider damping from viscous torques (proportional to angular velocity), an input deterministic torque, and random disturbances modeled by a Wiener process.
Finally, we develop an uncertainty propagation formula up to the second moment of the probability distribution on the group. 
The probability distribution constructed from the propagated mean and covariance fits the sample data better than the baseline extended Kalman filter.


\section{Background} \label{sec:back}

A \emph{group} is a set with a binary operation, called group operation, that obeys the group axioms \cite{chirikjian2011stochastic}. 
When a group is also an $N$-dimensional analytic manifold, it is called an $N$-dimensional \emph{Lie group}. 
Here, we only consider \emph{matrix Lie groups}, which are Lie groups whose elements can be represented by square matrices and where the group operation is matrix multiplication. 
The \emph{Lie algebra} $\mathcal{G}$ associated with a $N$-dimensional matrix Lie group $G$ is a $N$-dimensional vector space of matrices with a closed binary operation, $[X,Y]\doteq XY-YX, \  \text{where} \ X,Y\in \mathcal{G}$, called the Lie bracket.
The ``little ad'' operator is defined as $ad(X)Y\doteq [X,Y], \ \text{where} \ X,Y\in \mathcal{G}$.
Since the Lie algebra, $\mathcal{G}$, is a vector space, we can choose a set of basis $E_i\in \mathcal{G}$, $i=1,..., N$, so that any element $X\in\mathcal{G}$ can be written as $X = \sum_{i=1}^N x_iE_i$.
Using the basis, we define a ``$\vee$'' operation, $\vee:\mathcal{G}\rightarrow\mathbb{R}^N$, so that $X^\vee = \boldsymbol{x}=[x_1,...,x_N]^T\in\mathbb{R}^N$.
The inverse of the ``$\vee$'' is a ``$\wedge$'' so that $\boldsymbol{x}^\wedge = X$. 
The matrix representation of the ``little ad'' operator is then, $(ad(X)Y)^\vee = [ad(X)]Y^\vee$, where $[ad(X)]\in\mathbb{R}^{N\times N}$. 
Therefore,
\begin{equation}\label{eq:littlead}
[ad(X)] = \left[[X,E_1]^\vee,\cdots,[X,E_N]^\vee\right]
\end{equation}
where the $[\cdots]$ emphasizes that this is a matrix.

An element in the Lie algebra $\mathcal{G}$ can be converted into an element in the Lie group $G$ via matrix exponential:  $g(\boldsymbol{x})=\exp X \in G$, where $X\in\mathcal{G}$ and $\boldsymbol{x}=X^{\vee}$.
The inverse of the $\exp$ is the matrix logarithm denoted by $\log$.
Since the exponential map is bijective within a neighborhood of the identity, it can also be used to parameterize group elements: $g(\boldsymbol{x})\in G$, $\boldsymbol{x} =\log ^{\vee} (g)$.
The right/left Jacobian for this parameterization \cite{chirikjian2016harmonic} is defined by the matrix
\begin{equation}\label{eq:JR}
[J_r(g(\boldsymbol{x}))] = \left[\left(g^{-1}\frac{\partial g}{\partial x_1}\right)^\vee,\cdots,\left(g^{-1}\frac{\partial g}{\partial x_N}\right)^\vee\right],
\end{equation}
\begin{equation}\label{eq:JL}
[J_l(g(\boldsymbol{x}))] = \left[\left(\frac{\partial g}{\partial x_1}g^{-1}\right)^\vee,\cdots,\left(\frac{\partial g}{\partial x_N}g^{-1}\right)^\vee\right].
\end{equation} 
{It is not hard to show that $J_r(g(\boldsymbol{x}))\dot{\boldsymbol{x}}=(g^{-1}\dot{g})^{\vee}$.}
When $|{\det}\;J_l| \equiv |{\det}\;J_r|$, the group is called \emph{unimodular} and we can define a measure on the group: $dg=|\det J_l|d\boldsymbol{x}$.
Using this measure, the integration of a function on $G$ has the following invariant property:
\begin{equation*}
    \int_G f(g)dg=\!\int_G f(k\,\circ\, g)dg =\! \int_G f(g\,\circ\, k)dg =\! \int_G f(g^{-1})dg.
\end{equation*}
The right and left Lie derivatives of the group are defined for a differentiable function $f:G\rightarrow\mathbb{R}$ to be
\begin{equation}\label{eq:rightDerivativeNP}
    (E^r_i f)(g) \;\doteq\; \left.\frac{d}{dt}(f(g\circ e^{tE_i}))\right|_{t = 0},
\end{equation}
\begin{equation}\label{eq:leftDerivativeNP}
    (E^l_i f)(g) \;\doteq\; \left.\frac{d}{dt}(f(e^{-tE_i}\circ g))\right|_{t = 0},
\end{equation}
which is akin to the concept of a directional derivative on the group. 
Here $l$ and $r$ refer to which side of $g$ the perturbing exponential is applied.
Invariance under shifts are opposite to this.


\section{Problem Formulation}

Consider a three-dimensional rigid body with a moment of inertia tensor $I$ (in a body frame of reference). Its orientation is encoded by a three-dimensional rotation matrix $R\in SO(3)$. 
The body frame angular velocity of the body is defined as $\bm{\omega} = (R^T\dot{R})^\vee \in\mathbb{R}^3$ where $\vee$ is a bijection converting elements from the Lie algebra of $SO(3)$ to $\mathbb{R}^3$ (in other words, vectorizes $3\times 3$ skew-symmetric matrices). The angular momentum of the body is then $\bm{\ell} = I\bm{\omega}$.

This body is rotating subject to a deterministic torque $\boldsymbol{N}^*(t)$, a viscous torque that is proportional to the angular velocity,
 i.e. of the form $C\bm{\omega}_R$, as well as a random torque $\bm{\eta}$. 
The equation of motion is the familiar Euler equation for angular momentum with random torque:
\begin{equation}\label{eq:stochastic_l}
    \dot{\boldsymbol{\ell}} + (I^{-1}\boldsymbol{\ell})\times\boldsymbol{\ell} = -CI^{-1}\boldsymbol{\ell} + \bm{\eta} + \boldsymbol{N}^*.
\end{equation}
Writing this as a stochastic differential equation by noting that $\bm{\eta} dt = B' d\bm{W}$, where $d\bm{W}$ is a Wiener process increment with zero mean and variance $dt$, we have
\begin{equation}\label{eq:AngMomentSDE}
    d\boldsymbol{\ell} + (I^{-1}\boldsymbol{\ell})\times\boldsymbol{\ell}\;dt = -CI^{-1}\boldsymbol{\ell}\;dt + B'\;d\bm{W} + \boldsymbol{N}^*\,dt.
\end{equation}
Note that $B'$ controls the variance and `color' of the noise term. We also have the following equation
\begin{equation}\label{eq:RotSDE}
    (R^T\dot{R})^\vee dt = I^{-1}\bm{\ell}\;dt.
\end{equation}
The stochastic differential equations (\ref{eq:AngMomentSDE}) and (\ref{eq:RotSDE}) lead to a Fokker-Planck equation in phase space---the space of angular momentum and orientation---that can be solved to obtain the joint distribution $f(R,\boldsymbol{\ell},t)$. 

\subsection*{The group $SO(3)^T\ltimes\mathbb{R}^3$} From Hamiltonian dynamics, the phase space can be represented by the set of all orientation and angular momentum states that a rigid body can occupy. The space of rotations is $SO(3)$ and the space of angular momenta is $(\mathbb{R}^3,+)$, where $\mathbb{R}^3$ is imbued with the vector addition operation and seen as a group.

We imbue a semi-direct product structure to the phase space. From \cite{jayaraman2020black}, we see that this is in fact the (co-)tangent bundle group of $SO(3)$. 
A representative group element is 
\begin{equation}\label{eq:h_def}
    h(R,\bm{\ell}) = \left(
    \begin{array}{c|c}
    R^T & \bm{\ell}\\
    \hline
    \bm{0}^T & 1
    \end{array}\right),
\end{equation}
where $R\in SO(3)$ and $\boldsymbol{\ell}\in\mathbb{R}^3$.
Note that we parameterize rotations with $R^T$ instead of $R$ (as it would be for Euclidean motions), and we denote the group by $SO(3)^T\ltimes\mathbb{R}^3$. 
It is a unimodular group, whose Lie algebra, left/right Lie derivative, `$\vee$' operation, and ``little'' ad operator are derived in \cite{jayaraman2023inertial}. 
They will be used in the following derivation.
Note that the semi-direct product group structure has been exploited for the tangent bundle case in \cite{brockett1972tangent} and applied in estimation problems in \cite{ng2022equivariant,fornasier2022equivariant}. Here, we are using the cotangent bundle \cite{holm2011geometric}, which is different in the sense that the angular momentum is a co-tangent vector instead of a tangent vector of $SO(3)$.

We now observe that the left-hand side of the equations of motion (\ref{eq:AngMomentSDE}) and (\ref{eq:RotSDE}) can be written as $(\dot{h}h^{-1})^\vee dt = (dh\,h^{-1})^\vee$ where $h\in SO(3)^T\ltimes\mathbb{R}^3$ is given in (\ref{eq:h_def}). 
The key observation is
\begin{equation}
    \setlength{\arraycolsep}{1.5pt}
    \begin{aligned}
    (\dot{h}h^{-1})^\vee \! = \!
    \begin{pmatrix}
    \dot{R}^TR  && -(\dot{R}^TR)\boldsymbol{\ell} + \dot{\boldsymbol{\ell}}\\
    \boldsymbol{0}^T  &&  0
    \end{pmatrix}^\vee \!
    = \! \begin{pmatrix}
    \boldsymbol{\omega}\\
    \dot{\boldsymbol{\ell}} + \boldsymbol{\omega}\times\boldsymbol{\ell}
    \end{pmatrix}.
    \end{aligned}
\end{equation}
Comparing with (\ref{eq:AngMomentSDE}) and (\ref{eq:RotSDE}), we see that
\begin{equation}\label{eq:mainSDE}
    (\dot{h}h^{-1})^\vee dt = 
    \begin{pmatrix}
    I^{-1}\boldsymbol{\ell}\\
    -CI^{-1}\boldsymbol{\ell} + \boldsymbol{N}^*
    \end{pmatrix}\; dt +
    \begin{pmatrix}
    \mathbb{O} && \mathbb{O}\\
    \mathbb{O} && B'
    \end{pmatrix}\;d\boldsymbol{W},
\end{equation}
which is a (left) stochastic differential equation evolving on $SO(3)^T\ltimes\mathbb{R}^3$. Notably, the transposed rotations in $SO(3)^T\ltimes\mathbb{R}^3$ allow us to remove the troublesome cross-product term, thereby avoiding the need to deal with that term explicitly. Setting 
\begin{equation}\label{eq:m,B}
\boldsymbol{m} = 
\begin{pmatrix}
-I^{-1}\boldsymbol{\ell}\\
CI^{-1}\boldsymbol{\ell} - \boldsymbol{N}^*
\end{pmatrix}
\;\;\text{and}\;\;
B = 
\begin{pmatrix}
\mathbb{O} && \mathbb{O}\\
\mathbb{O} && B'
\end{pmatrix},
\end{equation}
the Fokker-Planck equation for the probability density function $u(h,t)$ on $SO(3)^T\ltimes\mathbb{R}^3$ is
\begin{equation}\label{eq:FPE_OU_Group}
    \frac{\partial u}{\partial t} =- E^l_i(m_i u) + \frac{1}{2} (BB^T)_{ij}E^l_iE^l_j u,
\end{equation}
using the result in \cite{chirikjian2011stochastic}, where $E^l_i$ is the left Lie derivative.
To make equations concise, we adopt the Einstein summation convention above, and it will be used throughout the paper.
 \label{sec:pro}


\section{Approximate Solution Methods for the Fokker-Planck equation} \label{sec:Approximate_FPE}

In this section, we propose an approximate method to extract the mean and covariance of the probability distribution $u(h,t)$ in (\ref{eq:FPE_OU_Group}). 
{
We first review the propagation step of the extended Kalman filter (EKF) on $SO(3)\times \mathbb{R}^3$ using the exponential coordinate and then introduce the Expansion of Moments (EOM) technique for $SO(3)^T\ltimes\mathbb{R}^3$.}

\subsection{Extended Kalman Filter (EKF)}
{
In this section, we derive the EKF propagation formulas for (\ref{eq:AngMomentSDE}) and (\ref{eq:RotSDE}) by parametrizing $SO(3)$ by the exponential map: $\boldsymbol{x}=\log^{\vee} R$ and $R(\boldsymbol{x})=\exp(\boldsymbol{x}^{\wedge})$.
The first step is to solve the deterministic trajectory,
\begin{equation}\label{eq:EKF_rot}
    \begin{cases} 
 \dot{\boldsymbol{x}}^* =J_r^{-1}(\boldsymbol{x}^*)I^{-1}\boldsymbol{\ell}^*\\
 \dot{\boldsymbol{\ell}}^* =-(I^{-1} \boldsymbol{\ell}^*)\times \boldsymbol{\ell}^*-CI^{-1}\boldsymbol{\ell}^*+\boldsymbol{N}^*,
\end{cases}
\end{equation}
where the first equation is the parametrized form of (\ref{eq:RotSDE}) and $J_r$ is the right Jacobian of $SO(3)$.
Then, we substitute the deterministic solution into the original SDE
\begin{equation}
    \begin{cases}
d(\boldsymbol{x}^*+\boldsymbol{\epsilon}_x)=J_r^{-1}(\boldsymbol{x}^*+\boldsymbol{\epsilon}_x)I^{-1}(\boldsymbol{\ell}^*+\boldsymbol{\epsilon} _l)dt\\
d(\boldsymbol{\ell}^*+\boldsymbol{\epsilon}_l)=[-(I^{-1}(\boldsymbol{\ell}^*+\boldsymbol{\epsilon}_l))\times (\boldsymbol{\ell}^*+\boldsymbol{\epsilon} _l)\\
\qquad \qquad \qquad  -CI^{-1}(\boldsymbol{\ell}^*+\boldsymbol{\epsilon}_l)+\boldsymbol{N}^*]dt+B'd\boldsymbol{W}
\end{cases}.
\end{equation}
Subtracting it by (\ref{eq:EKF_rot}) and keeping only the linear terms, we have
\begin{equation} \label{eq:EKF_linear}
\small
    \begin{aligned}
    d\begin{pmatrix} \boldsymbol{\epsilon}_x \\ \boldsymbol{\epsilon} _l \end{pmatrix}\!&\!=\!\begin{pmatrix} S(t) & J_r^{-1}(\boldsymbol{x}^*(t))I^{-1}  \\ 0 & -[I^{-1}\boldsymbol{\ell}^*(t)]^{\wedge}+{{\boldsymbol{\ell}^*}^{\wedge}(t)}I^{-1}-CI^{-1}  \end{pmatrix} \!\!\begin{pmatrix} \boldsymbol{\epsilon}_x \\ \boldsymbol{\epsilon} _l \end{pmatrix}dt\\&\qquad \qquad 
    +\begin{pmatrix} 0 \\ B' \end{pmatrix}d\boldsymbol{W}
    \end{aligned}
\end{equation}
where ${\boldsymbol{\ell}^*}^{\wedge}\doteq \ell^*_i \bar{E}_i$, $\bar{E}_i$ is the $i$th element of the basis of the Lie algebra of $SO(3)$, and
\begin{equation}
\begin{aligned}
        S(t)=\biggl[ \frac{\partial J^{-1}_r}{\partial x_1}\bigg|_{\boldsymbol{x}^*(t)}&I^{-1}\boldsymbol{\ell}^*(t),\frac{\partial J^{-1}_r}{\partial x_2}\bigg|_{\boldsymbol{x}^*(t)}I^{-1}\boldsymbol{\ell}^*(t) ,\\ &\frac{\partial J^{-1}_r}{\partial x_3}\bigg|_{\boldsymbol{x}^*(t)}I^{-1}\boldsymbol{\ell}^*(t)\biggr].
\end{aligned}
\end{equation}
Denote the mean of $[\boldsymbol{\epsilon}_x^T, \boldsymbol{\epsilon} _l^T]$ as $\boldsymbol{\mu}^T_{\epsilon}(t)$, the covariance matrix as $\Sigma(t)$, and the large coefficient matrix in (\ref{eq:EKF_linear}) as $A(t)$.
The propagation equations of mean and covariance are
\begin{equation} \label{eq:EKF_OU}
    \begin{cases}
 {\dot{\boldsymbol{\mu}}_{\epsilon}} = A\boldsymbol{\mu}_{\epsilon} \\
\dot{\Sigma} \,= A\Sigma+\Sigma A^T+BB^T
\end{cases}.
\end{equation}
where $B$ is defined in (\ref{eq:m,B})
. 
Since $\boldsymbol{\mu}_{\epsilon}(0)=\boldsymbol{0}$, (\ref{eq:EKF_OU}) indicates that $\boldsymbol{\mu}_{\epsilon}(t)=\boldsymbol{0}$ all the time.
So the mean and covariance matrix of EKF are $\boldsymbol{\mu}_{\text{EKF}}^T=[{{\boldsymbol{x}}^*}^T,\,{\boldsymbol{\ell}^*}^T]$ and $\Sigma_{\text{EKF}}=\Sigma$. 
}

{
Using the mean and covariance, we can construct a parametrized Gaussian solution of the form:
\begin{multline}\label{eq:EKF_ExpGaussSolution}
\tilde{u}_{\text{EKF}}(\boldsymbol{\xi},t) = \frac{1}{(2\pi)^3|\text{det}\;\Sigma_{\text{EKF}}(t)|^{\frac{1}{2}}}\times\\\exp\left(-\frac{1}{2}(\boldsymbol{\xi}-\boldsymbol{\mu}_{\text{EKF}}(t))^T \Sigma^{-1}_{\text{EKF}}(t)(\boldsymbol{\xi}-\boldsymbol{\mu}_{\text{EKF}}(t))\right),
\end{multline}
where here, $\boldsymbol{\xi}^T = [\boldsymbol{x}^T,\boldsymbol{\ell}^T]$ and $\boldsymbol{x}=\log^{\vee} R$. }

\subsection{Expansion of Moments (EOM)}
Here we utilize the chain rule and various identities of Lie group calculus \cite{chirikjian2011stochastic} in order to extract the group-theoretic mean and covariance from the Fokker-Planck equation in (\ref{eq:FPE_OU_Group}). This gives rise to a non-parametric coupled linear ordinary differential equation for mean and covariance.

We commence by defining the group theoretic mean and covariance as
\begin{equation}\label{eq:GroupMu}
\int_G \log^\vee(h \circ \mu^{-1}(t))\;u(h,t)\;dh \;\doteq\; \boldsymbol{0},
\end{equation}
and,
\begin{equation}\label{eq:GroupCov}
\Sigma(t) \;\doteq\; \int_G [\log^\vee(h \circ \mu^{-1}(t))][\log^\vee(h \circ \mu^{-1}(t))]^T\;u(h,t)\;dh.
\end{equation}
The position of $\mu(t)$ in the definition of mean and covariance differs from the position of the mean in the respective definitions in \cite{StochasticCart,wang2008nonparametric}, and the rationale for this choice will be made clear shortly.

We now proceed to write $\boldsymbol{m}$ from (\ref{eq:FPE_OU_Group}) as a function of $h\in SO(3)^T\ltimes\mathbb{R}^3$ by defining
\begin{equation}
    Q_G =
    \begin{pmatrix}
    -I^{-1}\\
    CI^{-1}
    \end{pmatrix}
    \begin{pmatrix}
    1 & 0 & 0 & 0\\
    0 & 1 & 0 & 0\\
    0 & 0 & 1 & 0
    \end{pmatrix}\,\,\text{and}\,\,
    \boldsymbol{\tau} = \begin{pmatrix}
    \boldsymbol{0}\\
    \boldsymbol{N}^*(t)
    \end{pmatrix}.
\end{equation}
A direct calculation yields $$m_i = [Q_G]_{ij}\boldsymbol{e}^T_j[h\cdot\boldsymbol{e}_4] - \tau_i,$$
since $[Q_G]$ is a $6\times 4$ matrix. Then we can write (\ref{eq:FPE_OU_Group}) as
\begin{multline}\label{eq:FPE_Group_U} 
    \hspace{-0.3cm}
    \frac{\partial u}{\partial t} = -[Q_G]_{ij}\boldsymbol{e}^T_jE^l_i(h u)\boldsymbol{e}_4 + E^l_i (\tau_i u) + \frac{1}{2}(BB^T)_{ij}E^l_iE^l_ju.
\end{multline}

In the following, it will be useful to express $u(h,t) = \rho(k(h,t),t)$ where $k = h\circ\mu^{-1}(t)$. Here, $\rho(k,t)$ can be interpreted as a distribution whose mean is at the identity.
By using the following equation:
\begin{equation}
    \mu(t+\varepsilon)=\exp[\varepsilon \cdot (\dot \mu  \mu^{-1})(t)+O(\varepsilon^2)]\circ \mu(t),
\end{equation}
we have
    \begin{align}  \label{eq:partial1}
    \! \frac{\partial u(h,t)}{\partial t} \!
    =& \! \lim_{\varepsilon\to 0} \frac{1}{\varepsilon}[\rho(k \circ \exp(-\varepsilon \! \cdot \! \dot \mu  \mu^{-1}) ,t)\! - \! \rho(k ,t) ] \! + \! \frac{\partial \rho(k,t)}{\partial t} \notag \\
    =& -(E^r_i\rho)\mathbf{e}_i^T(\dot \mu \mu^{-1})^{\vee} + \frac{\partial \rho(k,t)}{\partial t},
    \end{align}
where the definition of the right Lie derivative is used.

We can now explain the choice of positioning the $\mu(t)$ on the right of $h$ in $\log(h\circ\mu^{-1}(t))$ in (\ref{eq:GroupMu}), (\ref{eq:GroupCov}). Letting $u(h,t) = \rho(h\circ\mu^{-1}(t),t)$ instead of $u(h,t) = \rho(\mu^{-1}(t)\circ h,t)$ that was used in \cite{StochasticCart,park2010path,wang2006error}, we have the identity $(E^l_iu)(h,t) = (E^l_i\rho)(k,t)$ without additional terms. Substituting it and (\ref{eq:partial1}) into (\ref{eq:FPE_Group_U}), we obtain a Fokker-Planck equation in terms of $\rho$:
\begin{multline} \label{eq:FPE_Group_Rho}
    \frac{\partial \rho}{\partial t}=-[Q_G]_{ij}\boldsymbol{e}^T_jE^l_i(k\rho)\mu(t)\boldsymbol{e}_4 + E^l_i(\tau_i \rho)\\ + \frac{1}{2}(BB^T)_{ij}E^l_iE^l_j\rho +  (E^r_i\rho)\boldsymbol{e}_i^T(\dot{\mu}\mu^{-1})^\vee .
\end{multline}

In the following, we make use of the notation: $ad_i = [ad(E_i)]$, $ad_X = [ad(\log k)]$ and $ad_{[E_i,X]} = [ad(ad(E_i)\log k)]$. 
We also define
\begin{equation}\nonumber
    \Sigma' \doteq \begin{pmatrix}
    -\Sigma_{22} - \Sigma_{33} & \Sigma_{12} & \Sigma_{13}\\
    \Sigma_{12} & -\Sigma_{11}-\Sigma_{33} & \Sigma_{23}\\
    \Sigma_{13} & \Sigma_{23} & -\Sigma_{11}-\Sigma_{22}
    \end{pmatrix},
\end{equation}
and
\begin{equation}\nonumber
\begingroup 
\setlength\arraycolsep{1.5pt}
    \Sigma'' \doteq \begin{pmatrix}
    -2(\Sigma_{52}+\Sigma_{63}) & \Sigma_{42}+\Sigma_{51} & \Sigma_{43}+\Sigma_{61}\\
    \Sigma_{42}+\Sigma_{51} & -2(\Sigma_{41}+\Sigma_{63}) & \Sigma_{53}+\Sigma_{62}\\
    \Sigma_{43}+\Sigma_{61} & \Sigma_{53}+\Sigma_{62} & -2(\Sigma_{41}+\Sigma_{52})
    \end{pmatrix},
    \endgroup
\end{equation}
so that
\begin{align} \label{eq:A1_Sig}
    A_1(\Sigma) &\doteq \int_G ad_Xad_X\rho\,dk 
    = \left(\begin{array}{c|c}
    \Sigma' & \mathbb{O}\\
    \hline
    \Sigma'' & \Sigma'
    \end{array}\right),
\end{align}
and representing $\boldsymbol{\sigma}^T = [\Sigma_{62}-\Sigma_{35},\Sigma_{34}-\Sigma_{61},\Sigma_{51}-\Sigma_{42}]$,
\begin{align} \label{eq:A2_Sig}
    A_2(\Sigma) &\doteq \int_G X^2\,\rho\,dk = \left(\begin{array}{c|c}
    \Sigma' & \boldsymbol{\sigma}\\
    \hline
    \boldsymbol{0}^T & 0
    \end{array}\right).
\end{align}

The mean (\ref{eq:GroupMu}) and covariance (\ref{eq:GroupCov}) are now re-expressed in terms of $\rho(k,t)$ where $k = \exp X$,
\begin{equation}\label{eq:Final_GroupMean}
    \int_G [\log^\vee k] \rho(k,t)\,dk = \boldsymbol{0},
\end{equation}
\begin{equation}\label{eq:Final_GroupCov}
    \Sigma_{ij}=\int_G [\log^\vee k]_i[\log^\vee k]_j \rho(k,t)\,dk = \int_G x_ix_j \rho(k,t)\,dk. 
\end{equation}

Compared to EKF, where second-order terms are neglected, we make the assumption that third moments and higher of the function $\rho(k,t)$ are negligible. This gives a higher-order approximation and yields coupled ordinary differential equations for mean and covariance.

\subsubsection{Expression for $\mu(t)$}
Differentiating both sides of (\ref{eq:Final_GroupMean}) with respect to time and using (\ref{eq:FPE_Group_Rho}), we have:
\begin{theorem}
The equation for the evolution of the mean $\mu(t)$ is given implicitly by
\begin{equation}\label{eq:mu_eqn}
\boxed{
    \boldsymbol{G}_1(\Sigma,\mu) + \boldsymbol{G}_2(\Sigma) + \boldsymbol{G}_3(\Sigma) + \boldsymbol{G}_4(\Sigma,\mu,\dot{\mu}) = \boldsymbol{0}}
\end{equation}
where the vector-valued functions are
\begin{align}\nonumber
    \boldsymbol{G}_1(\Sigma,\mu) &= -[Q_G]_{ij}\int_G \boldsymbol{x}\boldsymbol{e}^T_jE^l_i(k \rho)\,dk\,\mu(t)\boldsymbol{e}_4\\\nonumber
    \boldsymbol{G}_2(\Sigma) &= \int_G\boldsymbol{x}E^l_i(\tau_i \rho)\,dk\\\nonumber
    \boldsymbol{G}_3(\Sigma) &= \frac{1}{2}\int_G\boldsymbol{x}(BB^T)_{ij}E^l_iE^l_j \rho\,dk\\\nonumber
    \boldsymbol{G}_4(\Sigma,\mu,\dot{\mu}) &= \int_G \boldsymbol{x}E^r_i\rho\,dk\,\boldsymbol{e}^T_i(\dot{\mu}\mu^{-1})^\vee.
\end{align}
\end{theorem}
In what follows, we proceed to simplify these integrals making use of the approximation $$k\approx \mathbb{I} + X + \frac{1}{2}X^2$$ as well as the approximations of the Lie derivatives in the appendix of \cite{jayaraman2023inertial}. 

In the evaluation of each of these integrals, we make use of integration of parts \cite{chirikjian2011stochastic} assuming that the function $\rho\in\mathcal{L}^2(SO(3)^T\ltimes\mathbb{R}^3)$, and all the integrals are convergent (i.e., the distribution decays to zero at `infinity' faster than the other functions that multiply with it in the integrand). Then, we have
\begin{equation}
    \int_G f_1 (E^{l/r}_i f_2)\,dk = -\int_G (E^{l/r}_i f_1) f_2\,dk,
\end{equation}
for `nice' functions $f_1(k)$ and $f_2(k)$, and the above holds for left and right Lie derivatives. 

Considering the first term, after using integration by parts and the above mentioned approximations, we see that
\begin{multline*}
    \boldsymbol{G}_1(\Sigma,\mu) \approx -[Q_G]_{ij}\int_G\left(\boldsymbol{e}_i\boldsymbol{e}_j^T + \frac{\boldsymbol{e}_i\boldsymbol{e}^T_j}{2}X^2 +\right.\\\left.+ \frac{1}{2}ad_i\boldsymbol{x}\boldsymbol{e}^T_jX + \frac{1}{12}ad_Xad_X\boldsymbol{e}_i\boldsymbol{e}_j^T\right)\,\rho\,dk\,\mu(t)\boldsymbol{e}_4,
\end{multline*}
assuming that third moments and higher can be neglected. Additionally, using $X =  [\log^\vee k]_m E_m = x_mE_m$, we have
\begin{align}\nonumber
    \int_G \boldsymbol{x}\boldsymbol{e}^T_jX\,\rho\,dk = \int_G \boldsymbol{x}x_m\,\rho\,dk\,\boldsymbol{e}^T_jE_m =  (\Sigma\boldsymbol{e}_m)\boldsymbol{e}^T_jE_m.
\end{align}
Taking together we have
\begin{equation}\label{eq:G1}
\begin{aligned}
    \boldsymbol{G}_1(\Sigma,\mu) \approx -\left\lbrack [Q_G]_{ij}\boldsymbol{e}_i\boldsymbol{e}^T_j + \frac{[Q_G]_{ij}}{2}\boldsymbol{e}_i\boldsymbol{e}^T_jA_1(\Sigma) \right.\\\left.+ \frac{[Q_G]_{ij}}{2}ad_i
    \Sigma\boldsymbol{e}_m\boldsymbol{e}^T_jE_m + \frac{[Q_G]_{ij}}{12}A_2(\Sigma)\boldsymbol{e}_i\boldsymbol{e}_j^T\right\rbrack\mu(t)\boldsymbol{e}_4.
    \end{aligned}
\end{equation}
We also see that
\begin{align} \label{eq:G2}
    \boldsymbol{G}_2(\Sigma) &\approx \int_G\,(\boldsymbol{e}_i + \frac{1}{12}ad_Xad_X\boldsymbol{e}_i)\,\tau_i\rho\,dk \notag \\
    &= \left(\mathbb{I} + \frac{1}{12}A_2(\Sigma)\right)\boldsymbol{\tau}.
\end{align}
Since
\begin{equation*}
    \boldsymbol{G}_3(\Sigma) = \frac{1}{2}\int_G(E^l_jE^l_i\boldsymbol{x})(BB^T)_{ij}\rho\,dk,
\end{equation*}
we can make use of the approximation formula for $E^l_i E^l_j \boldsymbol{x}$ in \cite{jayaraman2023inertial} in the simplification process. Noting that
\begin{equation}
    [E_i,E_j] = C^k_{ij}E_k,
\end{equation}
where $C^k_{ij}$ are the structure constants and we can write
\begin{align}
    ad_{[E_j,X]}ad_X\boldsymbol{e}_i = [[E_j,X],[X,E_i]]^\vee 
    = x_mx_n C^{k}_{mi}C^l_{nj}C^p_{kl}\boldsymbol{e}_p,
\end{align}
and similarly
\begin{equation}
    ad_Xad_{[E_j,X]}\boldsymbol{e}_i = x_mx_n C^p_{ml}C^k_{jn}C^l_{ki}\boldsymbol{e}_p.
\end{equation}
Putting them all together we have
\begin{equation}\label{eq:G3}
\begin{aligned}
    \boldsymbol{G}_3(\Sigma) \approx (BB^T)_{ij}\left(\frac{1}{4}ad_i\boldsymbol{e}_j + \frac{1}{48}ad_iA_2(\Sigma)\boldsymbol{e}_j\right) \\+ \frac{1}{48}(BB^T)_{ij}\Sigma_{mn}(C^k_{mi}C^l_{nj}C^p_{kl} + C^p_{ml}C^k_{jn}C^l_{ki})\boldsymbol{e}_p.\end{aligned}
\end{equation}
Finally, similar to (\ref{eq:G2}),
\begin{align}
    \boldsymbol{G}_4(\Sigma,\mu,\dot{\mu}) 
    \label{eq:G4}
    &\approx -\left(\mathbb{I} + \frac{1}{12}A_2(\Sigma)\right)(\dot{\mu}\mu^{-1})^\vee.
\end{align}
These results can be substituted in (\ref{eq:mu_eqn}) to obtain a nonlinear ODE in $\mu$. This ODE also involves $\Sigma$, and we require an additional equation to close the system.

\subsubsection{Expression for $\Sigma(t)$} Differentiating both sides of (\ref{eq:Final_GroupCov}) with respect to time and using (\ref{eq:FPE_Group_Rho}), we have:
\begin{theorem}
The equation for the evolution of the covariance $\Sigma(t)$ is given as
\begin{equation}\label{eq:sigma_eqn}
    \boxed{\dot{\Sigma} = F_1(\Sigma,\mu) + F_2(\Sigma) + F_3(\Sigma) + F_4(\Sigma,\mu,\dot{\mu})}
\end{equation}
where
\begin{align*}
    F_1(\Sigma,\mu) &= - \int_G \boldsymbol{x}\boldsymbol{x}^T [Q_G]_{ij}\boldsymbol{e}^T_jE^l_i(k\mu(t)\rho)\boldsymbol{e}_4\,dk,\\
    F_2(\Sigma) &= \int_G \boldsymbol{x}\boldsymbol{x}^T E^l_i(\tau_i\rho)\,dk,\\
    F_3(\Sigma) &= \frac{1}{2}\int_G \boldsymbol{x}\boldsymbol{x}^T (BB^T)_{ij}E^l_iE^l_j\rho\,dk,\\
    F_4(\Sigma,\mu,\dot{\mu}) &=\left[ \int_G\boldsymbol{x}\boldsymbol{x}^T E^r_i\rho\,dk\right] \boldsymbol{e}^T_i(\dot{\mu}\mu^{-1})^\vee.
\end{align*}
\end{theorem}

We shall attempt to simplify these integrals to second order in $X$ using the approximation
\begin{equation}\nonumber
    k  = \exp X \approx \mathbb{I} + X,
\end{equation}
and by making use of integration by parts and the results in the appendix of \cite{jayaraman2023inertial}. For instance, 
\begin{multline}\label{eq:Q_RHS}
    F_1(\Sigma,\mu) \approx  -[Q_G]_{ij} \int_G\left(\boldsymbol{x}\boldsymbol{e}^T_i + \boldsymbol{e}_i\boldsymbol{x}^T + \frac{1}{2}\boldsymbol{x}\boldsymbol{x}^T ad_i^T \right.\\\left.+\frac{1}{2} ad_i\boldsymbol{x}\boldsymbol{x}^T \right) \boldsymbol{e}^T_j k\rho \,dk\,\mu(t)\boldsymbol{e}_4.
\end{multline}
Substituting the approximation $k\approx \mathbb{I} + X$ and the definition of the mean into (\ref{eq:Q_RHS}) we get, correct to the second moment of $\rho(k,t)$,
\begin{multline}\label{eq:Q_int_simplified_1}
F_1(\Sigma,\mu) \approx -[Q_G]_{ij}\left[\int_G (\boldsymbol{x}\boldsymbol{e}_i^T+\boldsymbol{e}_i \boldsymbol{x}^T)\boldsymbol{e}_j^TX\mu(t)\boldsymbol{e}_4 \rho dk   \right.\\ \left.+ \frac{1}{2}\int_G \left(\boldsymbol{x}\boldsymbol{x}^Tad_i^T + ad_i\boldsymbol{x}\boldsymbol{x}^T\right) \boldsymbol{e}_j^T\mu(t)\boldsymbol{e}_4\rho dk \right],
\end{multline}
and we can expand $\boldsymbol{e}_j^TX\mu(t)\boldsymbol{e}_4 =  x_m\boldsymbol{e}_j^TE_m\mu(t)\boldsymbol{e}_4$. We remind the reader that $\boldsymbol{e}_j^T(\log k)\mu(t)\boldsymbol{e}_4$, $\boldsymbol{e}^T_jE_m\mu(t)\boldsymbol{e}_4 $ and $\boldsymbol{e}^T_j\mu(t)\boldsymbol{e}_4$ are scalars. Thus, we can write
(\ref{eq:Q_int_simplified_1}) as
\begin{equation}\label{eq:F1}
\begin{aligned}
F_1(\Sigma,\mu)\approx -[Q_G]_{ij} \left[\frac{\boldsymbol{e}_j^T\mu(t)\boldsymbol{e}_4}{2}\left(\Sigma ad_i^T + ad_i\Sigma\right)\right.\\\left.+\sum_{m=1}^6(\boldsymbol{e}_j^TE_m\mu(t)\boldsymbol{e}_4) [(\Sigma\boldsymbol{e}_m)\boldsymbol{e}^T_i + \boldsymbol{e}_i(\Sigma\boldsymbol{e}_m)^T]\right].
    \end{aligned}
\end{equation}

To second order, we can express $F_2(\Sigma)$ as
\begin{equation}\label{eq:F2}
    F_2(\Sigma) \approx \frac{\tau_i}{2}\left(\Sigma ad_i^T + ad_i\Sigma\right)
\end{equation}
and $F_3(\Sigma)$ as
\begin{multline}
    \nonumber
    F_3(\Sigma) \approx\frac{1}{2} \int_G (I_P(i,j) + I_P(i,j)^T + I_S(i,j) \\+ I_S(i,j)^T)(BB^T)_{ij} \rho\,dk,
\end{multline}
where
$$
I_P(i,j) = \frac{1}{4}ad_iad_j\boldsymbol{x}\boldsymbol{x}^T + \frac{1}{12}ad_jad_X\boldsymbol{e}_i\boldsymbol{x}^T + \frac{1}{12}ad_Xad_j\boldsymbol{e}_i\boldsymbol{x}^T
$$ and $$ \begin{aligned} I_S(i,j) = &\boldsymbol{e}_j\boldsymbol{e}_i^T + \frac{1}{4}ad_j\boldsymbol{x}\boldsymbol{x}^Tad_i^T + \frac{1}{12}\boldsymbol{e}_j\boldsymbol{e}^T_iad_X^Tad_X^T + \\ &\frac{1}{12}ad_Xad_X\boldsymbol{e}_j\boldsymbol{e}^T_i. \end{aligned}$$
Now,
\begin{align*}
ad_jad_X\boldsymbol{e}_i\boldsymbol{x}^T &= [E_j,[X, E_i]]^\vee\boldsymbol{x}^T\\ 
&= -[E_j,[E_i,X]]^\vee\boldsymbol{x}^T\\ &= -ad_jad_i\boldsymbol{x}\boldsymbol{x}^T,
\end{align*}
and similarly,
\begin{align*}
ad_Xad_j\boldsymbol{e}_i\boldsymbol{x}^T 
&= -ad_{[E_j,E_i]}\boldsymbol{x}\boldsymbol{x}^T.
\end{align*}
Then we have
\begin{equation}\label{eq:F3}
\begin{aligned}
    F_3(\Sigma) 
    \approx (BB^T) + \frac{1}{8} (BB^T)_{ij}\left([I^{\text{I}}_{ij}(\Sigma)] + [I^{\text{II}}_{ij}(\Sigma)]\right) + &\\ \frac{1}{24} (BB^T)_{ij}\left([I^{\text{III}}_{ij}(\Sigma)] + [I^{\text{IV}}_{ij}(\Sigma)] + [I^{\text{V}}_{ij}(\Sigma)] + [I^{\text{VI}}_{ij}(\Sigma)]\right)&,
\end{aligned}
\end{equation}
where
\begin{align*}
    I^{\text{I}}_{ij}(\Sigma) &= [ad_iad_j\Sigma] + [ad_iad_j\Sigma]^T\\
    I^{\text{II}}_{ij}(\Sigma) &= [ad_i\Sigma ad_j^T] + [ad_i\Sigma ad_j^T]^T\\
    I^{\text{III}}_{ij}(\Sigma) &= \boldsymbol{e}_j\boldsymbol{e}^T_iA^T_2(\Sigma) + (\boldsymbol{e}_j\boldsymbol{e}^T_iA^T_2(\Sigma))^T\\
    I^{\text{IV}}_{ij}(\Sigma) &=  A_2(\Sigma)\boldsymbol{e}_j\boldsymbol{e}^T_i+ (A_2(\Sigma)\boldsymbol{e}_j\boldsymbol{e}^T_i)^T\\
    I^{\text{V}}_{ij}(\Sigma) &= -ad_jad_i\Sigma-(ad_jad_i\Sigma)^T\\
    I^{\text{VI}}_{ij}(\Sigma) &= -ad_{[E_j,E_i]}\Sigma-(ad_{[E_j,E_i]}\Sigma)^T.
\end{align*}

$F_4$ can be approximated as
\begin{equation}\label{eq:F4}
    F_4(\Sigma,\mu,\dot{\mu}) \approx \frac{1}{2}\left(\Sigma ad_j^T + ad_j\Sigma\right)\boldsymbol{e}^T_j(\dot{\mu}\mu^{-1})^\vee.
\end{equation}

Combining the expressions for $F_1(\Sigma,\mu)$ in (\ref{eq:F1}), $F_2(\Sigma)$ in (\ref{eq:F2}), $F_3(\Sigma)$ in (\ref{eq:F3}) and $F_4(\Sigma,\mu,\dot{\mu})$ in (\ref{eq:F4}), we see that we can express the right hand side of the equation (\ref{eq:sigma_eqn}) in terms of $\Sigma$, thus providing matrix ODEs that we can use for propagation.

Solving (\ref{eq:sigma_eqn}) together with (\ref{eq:mu_eqn}) numerically yields an approximation for the mean and covariance over time. To summarize, the approximation is valid to the second moments of the distribution, since when performing the chain rule and integration by parts all cubic and higher powers of $X = [\log k]$ were neglected.  

Using these results for $\mu(t)$ and $\Sigma(t)$, and using the subscript EOM to distinguish from the EKF mean/covariance, we can cast the solution as a Gaussian on the exponential coordinate of $SO(3)^T\ltimes\mathbb{R}^3$:
\begin{multline}\label{eq:Solution_EOM}
\hspace{-0.4cm}
\tilde{u}_{\text{EOM}}(\boldsymbol{y},t)\!=\! \frac{1}{(2\pi)^3|\text{det}\,\Sigma_{\text{EOM}}(t)|^{\frac{1}{2}}}  \exp\!\left(\!-\frac{1}{2}[\boldsymbol{y}^T\Sigma_{\text{EOM}}^{-1}(t)\boldsymbol{y}]\!\right),
\end{multline}
where $\boldsymbol{y} = \log^\vee (h\circ\mu_{\text{EOM}}^{-1}(t))$. 
This distribution is different from the  Gaussians that have been introduced previously in \cite{wang2006error,StochasticCart,jayaraman2020black,sun2021lie}, where the distributions are defined on the group, instead of on the exponential coordinate.
The following equation converts (\ref{eq:Solution_EOM}) back to the probability density function $u_{\text{EOM}}(h,t)$ on the group:
\begin{equation}
\begin{aligned}
u_{\text{EOM}}(h,t)dh=u_{\text{EOM}}(h(\boldsymbol{y}),t)|J_l (h(\boldsymbol{y}))|d\boldsymbol{y}
=\tilde{u}_{\text{EOM}}(\boldsymbol{y},t)d\boldsymbol{y}.
\end{aligned}
\end{equation}


\section{Numerical Results} \label{sec:Results}

Consider a body fixed frame that is oriented along the eigenvectors of the moment of inertia tensor $I$.
In this frame, we have a diagonal moment of inertia tensor, $I = \text{diag}(I_1,I_2,I_3)$. Then assume that the viscous tensor is of the form $C = c\mathbb{I}$ (i.e., diagonal and isotropic); likewise, assume that ${B}$ is given as
\begin{equation}
    {B} = 
    \begin{pmatrix}
    \mathbb{O} && \mathbb{O}\\
    \mathbb{O} && b\mathbb{I}
    \end{pmatrix},
\end{equation}
where we choose these forms to keep the model simple (although the equations (\ref{eq:mu_eqn}) and (\ref{eq:sigma_eqn}) can handle more general cases). In the results that follow, we choose $c = 1$ and $b = 1$.

Consider the moment of inertia tensor of the form
\begin{equation*}
    I = 
    \begin{pmatrix}
    2.070 && 0 && 0\\
    0 && 1.532 && 0\\
    0 && 0 && 1.236
    \end{pmatrix},
\end{equation*}
where the relative values are based on the inertia matrix considered in \cite{fujikawa1995spacecraft}. The fact that $I$ is not a multiple of the diagonal matrix ensures that the nonlinear cross-product term in the equation (\ref{eq:AngMomentSDE}) does not vanish. 

We consider two different deterministic angular momentum trajectories:
\begin{equation} \label{eq:det_l}
    \boldsymbol{\ell}^*_1(t) = 
    \begin{pmatrix}
    0\\
    t\\
    2t
    \end{pmatrix}\,\,\text{and}\,\,
    \boldsymbol{\ell}^*_2(t) = 
    \begin{pmatrix}
    0\\
    t+1\\
    2t+1
    \end{pmatrix}.
\end{equation}
For both prescribed angular momentum trajectories, the deterministic torque $\boldsymbol{N}^*(t)$ is obtained through a numerical evaluation of (\ref{eq:stochastic_l}) without introducing noise. 

\subsection{Sampling \& Numerical Scheme} \label{sec:A}
We simulate (\ref{eq:AngMomentSDE}) using a modified improved Euler's scheme for stochastic differential equations \cite{roberts2012modify}.
The trajectory update on rotations (\ref{eq:RotSDE}) is modified to be
\begin{equation}
    R_i(t + dt) = R_i(t) \circ \exp \left(\frac{dt}{2}(I^{-1}\boldsymbol{\ell}(t)+I^{-1}\boldsymbol{\ell}(t+dt))^{\wedge} \right).
\end{equation}
Using the representation of the group element in (\ref{eq:h_def}), we have $h_i(t)=h(R_i(t),\boldsymbol{\ell}_i(t))$, which defines the trajectory on the cotangent bundle group for a single particle $i$.
In total, we sample $N = 5,000,000$ particles over a time of 1 unit. 
A time step of $10^{-3}$ units is used.

We employ an iterative method to calculate the sample mean. The mean is initiated at
\begin{equation}
    \mu^0_{s}(t) = \exp\left(\frac{1}{N}\sum_{i=1}^N \log h_i(t) \right).
\end{equation}
Then this estimate is refined until the error defined by $||\frac{1}{N}\sum_{i=1}^N \log(h_i(t)\circ [\mu^{j-1}_s]^{-1}(t))||_F$ reduces to be less than $10^{-6}$. 
The rule for an iterative update of the mean is given by
\begin{equation} \label{eq:mean_iter}
    \mu^{j}_{s}(t) = \exp\left(\frac{1}{N}\sum_{i=1}^N \log(h_i(t)\circ [\mu^{j-1}]^{-1}_s(t))\right)\circ\mu^{j-1}_s(t).
\end{equation}
The result is the group-theoretic mean that can be compared with the mean obtained from the EOM method.
For the EKF method which estimates the mean on the direct product group $SO(3)\times\mathbb{R}^3$, we use (\ref{eq:mean_iter}) to obtain the sample mean for rotation and calculate the arithmetic mean for angular momentum for comparison.

To solve the ordinary differential equations of EKF/EOM, we use the improved Euler's method with the same time step. 
The implicit equation for the evolution of mean (\ref{eq:mu_eqn}) is made explicit by inverting the coefficient matrix in (\ref{eq:G4}).
When the time is short, the covariance is small and thus the matrix is invertible.

\subsection{Evaluation Metric}
We evaluate the mean of the rotation and angular momentum separately. 
A Frobenius error can be defined as
\begin{equation}
    \setlength{\abovedisplayskip}{2pt}
    \setlength{\belowdisplayskip}{2pt}
    \text{Error} = {||q_s(t) - q_{p}(t)||_F},
\end{equation}
where $||\cdots||_F$ is the Frobenius norm, $q$ is the mean for either rotation or angular momentum, and the subscript $s, p$ stands for `sample' and `propagated' respectively. 
Note that the sample means for EKF and EOM are different as stated in \ref{sec:A}.

{
We evaluate the Gaussian distributions constructed from the propagated mean and covariance by computing the negative log-likelihoods for both the EKF and EOM models.
The expression of the negative log-likelihood for Gaussians in (\ref{eq:EKF_ExpGaussSolution}) and (\ref{eq:Solution_EOM}) is
\begin{equation}
    \mathcal{L}_{method}= \log\left( (2\pi)^3|\det \Sigma|^{1/2} \right)+\frac{1}{2N}\sum_{i=1}^{N_s}\boldsymbol{z}_i^T \Sigma^{-1}\boldsymbol{z}_i,
\end{equation}
where we have divided the traditional negative log-likelihood by $N$ to avoid a large value.
In (\ref{eq:EKF_ExpGaussSolution}), $\boldsymbol{z}=\boldsymbol{\xi}-\boldsymbol{\mu}_{\text{EKF}}$, and in (\ref{eq:Solution_EOM}), $\boldsymbol{z}=\log^{\vee}(h\circ \mu_{\text{EOM}}^{-1})$.
We plot the difference of negative log-likelihood between our method EOM and the baseline method EKF, \textit{i}. \textit{e}. $\mathcal{L}_{\text{EOM}}-\mathcal{L}_{\text{EKF}}$. 
The more negative this difference, the better the fit from EOM as opposed to EKF.
}

\subsection{Experiments}
\begin{figure}[thpb]
      \setlength{\abovecaptionskip}{0.cm}
      \centering
      \includegraphics[width=0.47\textwidth]{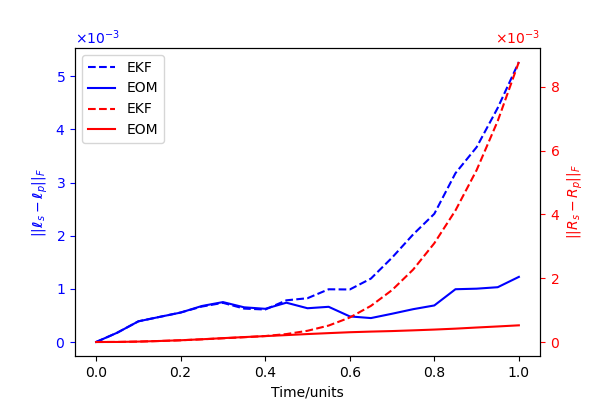}
      \caption{The evolution of the absolute error of angular momentum mean (blue) and rotation mean (red) for Trajectory 1.}
      \vspace{0cm}
      \label{fig:MuErrorTraj1_EKF_EoM}
\end{figure}
\begin{figure}[thpb]
      \setlength{\abovecaptionskip}{0.cm}
      \setlength{\belowcaptionskip}{0.cm}
      \centering
      \includegraphics[width=0.43\textwidth]{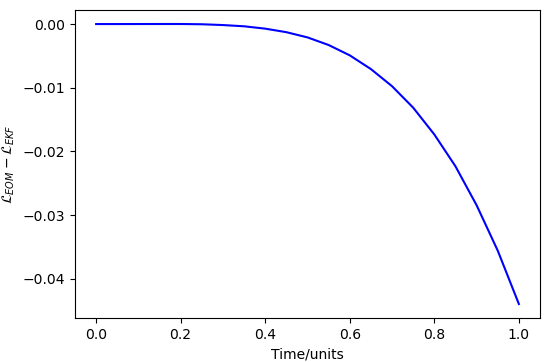}
      \caption{The evolution of the difference of negative log-likelihood for Trajectory 1.}
      \label{fig:PP_Traj1_EKF_EoM}
\end{figure}

In this section, we compare EKF/EOM using the stochastic sample paths for the deterministic angular momentum trajectories in (\ref{eq:det_l}).

\begin{figure}[thpb]
      \setlength{\abovecaptionskip}{0.cm}
      \centering
      \includegraphics[width=0.48\textwidth]{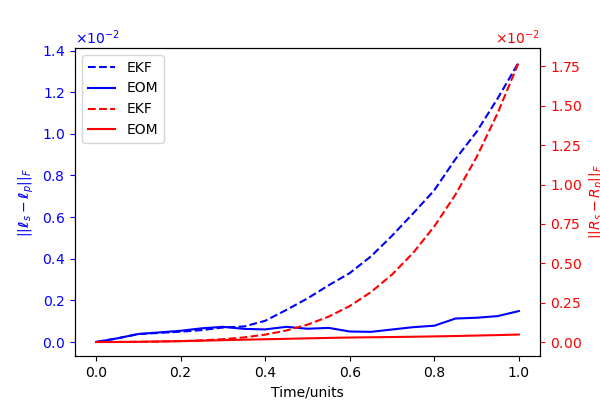}
      \caption{The evolution of the absolute error of angular momentum mean (blue) and rotation mean (red) for Trajectory 2.}
      \vspace{0cm}
      \label{fig:MuErrorTraj2_EKF_EoM}
\end{figure}
\begin{figure}[thpb]
      \setlength{\abovecaptionskip}{0.cm}
      \centering
      \includegraphics[width=0.44\textwidth]{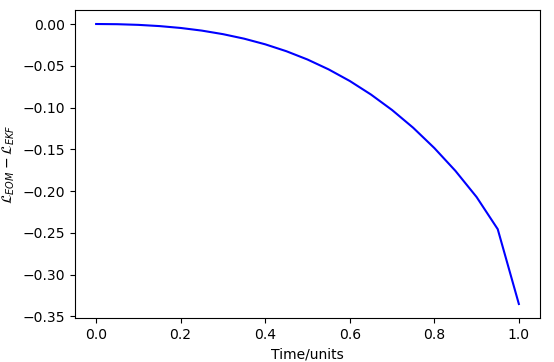}
      \caption{The evolution of the difference of negative log-likelihood for Trajectory 2.}
      \label{fig:PP_Traj2_EKF_EoM}
\end{figure}

We see from Figures \ref{fig:MuErrorTraj1_EKF_EoM} and \ref{fig:MuErrorTraj2_EKF_EoM} that for large time steps, the mean estimation with EOM is better than that by using EKF. Also from Figures \ref{fig:PP_Traj1_EKF_EoM} and \ref{fig:PP_Traj2_EKF_EoM}, the probability distribution constructed from the EOM mean and covariance fits the sample data better than that from EKF for large time steps. 

At sufficiently large times, we would expect both EKF and EOM to fail, due to the breakdown of the Gaussian assumption on the Lie group. 
It is more practical to take a new measurement before these large times are reached, in the context of the state estimation application.


\section{Conclusions \& Future Work}\label{sec:Concl}

The paper develops a joint rotation and angular momentum uncertainty propagation theory for dynamic state estimation problems. 
The mean and covariance propagation equations for forced rotational Brownian motion are derived non-parametrically from the Fokker-Planck equation on the group $SO(3)^T\ltimes\mathbb{R}^3$.
The equations are then approximated up to the second moment in the probability distribution function. 
Experiments show that the resulting distribution fits the data better than EKF.

For future work, the resulting probability density $u(h(R,\boldsymbol{\ell}),t)$ can be combined with a sensor model to construct a filter for use in dynamic state estimation. 
Additionally, since the propagated results are only valid for small times/covariances, it is important to extend the applicability of the theory to larger covariances, for instance by using the Fourier transform for $SE(3)$ to simplify the governing Fokker-Planck equation in (\ref{eq:FPE_OU_Group}).


\section{Acknowledgements}

The authors would like to thank Professor Domenico Campolo for introducing the authors to the cotangent bundle structure and Professor Victor Solo for insightful discussions on the topic.


\bibliography{References_RAL}

\end{document}